\RequirePackage{fix-cm}
\documentclass[
aps,prl,
amsmath,amssymb,superscriptaddress,
preprint
]{revtex4-1}
\usepackage{graphicx}
\usepackage{dcolumn}
\usepackage{bm}
\usepackage{color}
\usepackage[utf8]{inputenc}
\usepackage[T1]{fontenc}
\usepackage{mathptmx}

\begin{document}

\title{
Deep vacancy induced low-density fluxional interfacial water
}

\author{Keyang Liu}
\affiliation{School of Physics, Peking University, Beijing 100871, People’s Republic of China}
\author{Jianqing Guo}
\affiliation{International Center for Quantum Materials, School of Physics, Peking University, Beijing 100871, People’s Republic of China}
\author{Weizhong Fu}
\affiliation{School of Physics, Peking University, Beijing 100871, People’s Republic of China}
\author{Ji Chen}
\email{ji.chen@pku.edu.cn}
\affiliation
{School of Physics, Peking University, Beijing 100871, People’s Republic of China}
\affiliation
{ Collaborative Innovation Center of Quantum Matter, Beijing 100871, People’s Republic of China}



\begin{abstract}
Interfacial water on transition metal oxides such as TiO$_2$ has been
widely studied because of their structural complexity and scientific relevance in e.g. photocatalysis and ice growth. 
Using \textit{ab initio} molecular dynamics, we find 
that interfacial water on anatase (101) surface features an unconventional fluxional structure with reduced contact layer density.
The density reduction and flexibility of interfacial water are induced by oxygen vacancy defects deeply located in the subsurface. 
Our study proposes a fresh perspective of the anatase/water interface, raising the importance of non-trivial long-range effects caused by deep defects. 
These often-neglected effects highlight the necessity and challenges of the state-of-the-art simulation and experimental probing of solid liquid interfaces.
\end{abstract}

\date{\today}

\maketitle
Water is one of the most ubiquitous materials in nature and has an incredibly diverse range of structures and properties at the interfaces with other materials\cite{hodgsonWaterAdsorptionWetting2009a,bjorneholm_water_2016,debenedetti_chemical_2017}. 
Aqueous interfaces are fundamental to many natural phenomena and modern technologies\cite{kuhlenbeckWellOrderedTransitionMetal2013,sossoCrystalNucleationLiquids2016a,striolo_carbon-water_2016}.
Understanding the molecular structure of interfacial liquid water is key to examining the physical and chemical processes at interfaces. 
For example, ordering and density of interfacial water are two key elements for heterogeneous ice crystallization \cite{sossoCrystalNucleationLiquids2016a,fitzner_predicting_2020}; 
electric double layers formed at water solid interfaces control many electrochemical reactions \cite{sulpizi_silicawater_2012,kattirtzi_microscopic_2017,magnussen_toward_2019,zhang_coupling_2019};
interfacial water crucially affects energy converting in modern nanotechnologies \cite{deangelisTheoreticalStudiesAnatase2014,selcukFacetdependentTrappingDynamics2016,rao_solar_2017,tocci_ab_2020}. 
Because of these important applications, many experimental studies have been carried out to establish microscopic structure of interfacial water on surfaces, including e.g. high resolution imaging in vacuum \cite{stipe_single-molecule_1998}, in-situ diffraction measurements \cite{henderson_interaction_2002} and vibrational spectroscopy \cite{shen_sum_frequency_2006,li_situ_2019}. Molecular dynamics simulations based on density functional theory and force fields have also been very successful in elucidating interfacial structure and water dissociation \cite{carrascoMolecularPerspectiveWater2012,cheng_redox_2014,bjorneholm_water_2016,andrade_free_2020}.

From a different perspective, understanding the details of substrate materials can lead to fine tuning of properties at solid liquid interfaces \cite{bjorneholm_water_2016}. 
For example, in titanium dioxide, a material that has been widely investigated for photo-catalysis, reduced defects such as oxygen vacancies are considered to be ubiquitous \cite{pang_structure_2013,diebold_perspective_2017}. 
Studies have shown that defects on surface can react with water and subsurface defects can migrate to surface to interact with water \cite{bikondoaDirectVisualizationDefectmediated2006,liInterplayWaterTiO2014,deskins_observation_2020}. 
Besides, excess electrons induced by defects can also migrate and affect water adsorption and dissociation \cite{deskins_defining_2010,selcukFacetdependentTrappingDynamics2016,yimVisualizationWaterInducedSurface2018,chenSmallPolaronsJanus2020}. 
These interfacial effects, involving direct chemical interactions between water and surface, are intuitively understandable \cite{diebold_perspective_2017}. 
In contrast, long range physical effects induced by defects are generally less important.
However, the fact that there is a delicate balance between water adsorption and hydrogen bonding network suggests interfacial water properties may be sensitive to deep defects in subsurface as well.

In this study, we investigate the sensitivity of interfacial water structure to oxygen vacancy, a natural defect in TiO$_{2}$. We find that oxygen vacancies, surprisingly even very deep vacancies, have strong impacts on the structures of interfacial water. The impacts feature a density reduction and a fluxional behavior of interfacial water. These effects are not a direct consequence of reactions with surface oxygen vacancies and polarons, but rather due to a delicate balance of (i) hydrogen bonding between water molecules, (ii) adsorption of water, (iii) hydrogen bonding interactions between water and the substrate, and (iv) long range electrostatic interactions.
 
\textit{Ab initio} molecular dynamics (AIMD) simulations \cite{car_unified_1985} were carried out using density functional theory (DFT) with the Vienna \textit{ab initio} simulation package(VASP) \cite{kresse_efficient_1996}. 
The Van der Waals inclusive optB86-vdW exchange correlation functional was employed \cite{klimes_chemical_2010,klimes_van_2011}. 
Additional simulations with SCAN-rvv10 exchange correlation functional\cite{PhysRevLett.115.036402,PhysRevX.6.041005} and with Hubbard corrections (U=3.9 eV)\cite{dudarev_electron-energy-loss_1998} were performed to test the validity of conclusions.
Electron–ion interactions were described using the projector augmented wave method, with Ti(3d, 4s), O(2s, 2p), and H(1s) electrons treated explicitly as valence electrons \cite{kresse_ultrasoft_1999}. 
Wave functions were expanded in plane-waves up to a kinetic energy cutoff of 500 eV. K-point sampling was restricted to the $\Gamma$ point only. The deuterium mass (2 amu) was used for hydrogen to allow for a larger time step of 1 fs. Temperature was controlled using Nos$\acute{e}$-Hoover thermostat at 330 K within NVT ensemble.
Ten different configurations were considered for anatase(101), including a stoichiometric substrate and nine substrates each containing one oxygen vacancy (V$_O$) located at different depths. The oxygen vacancy concentration in our supercell is $\frac{1}{120}$, which is within the typical range in reduced TiO$_{2}$. Taking the stoichiometric configuration as an example, a periodically repeated $1\times3$ supercell of size $10.21{\AA}\times11.328{\AA}\times31.58{\AA}$ was used in our simulation, corresponding to a composition of ten Ti-layers and twenty O-layers ($Ti_{60}O_{120}$) and a (101) surface area of $10.21\times11.328$ \AA$^2$. Water slabs including 52 water molecules were constructed with the bulk water density (1.02 g/cm$^3$). After reaching equilibrium in e.g. 10 ps, trajectories were collected for up to 120 ps. The pressure is within 50 MPa throughout the simulations. We carry out additional simulations with 15 \AA\ vacuum above 76 water molecules and a substrate with 36 Ti atoms and 72/71 oxygen atoms (V$_{O}$8/pristine system). The bottom layer of substrate is fixed to keep the bulk TiO$_{2}$ properties.(Fig. S15)
We also tested other models, including anatase(001) (Ti$_{90}$O$_{180}$), rutile(110) (Ti$_{64}$O$_{128}$) and silica(0001) (Si$_{24}$O$_{48}$). Further details and convergence tests are described in the supporting information.

\begin{figure}
\includegraphics[width=3.25 in]{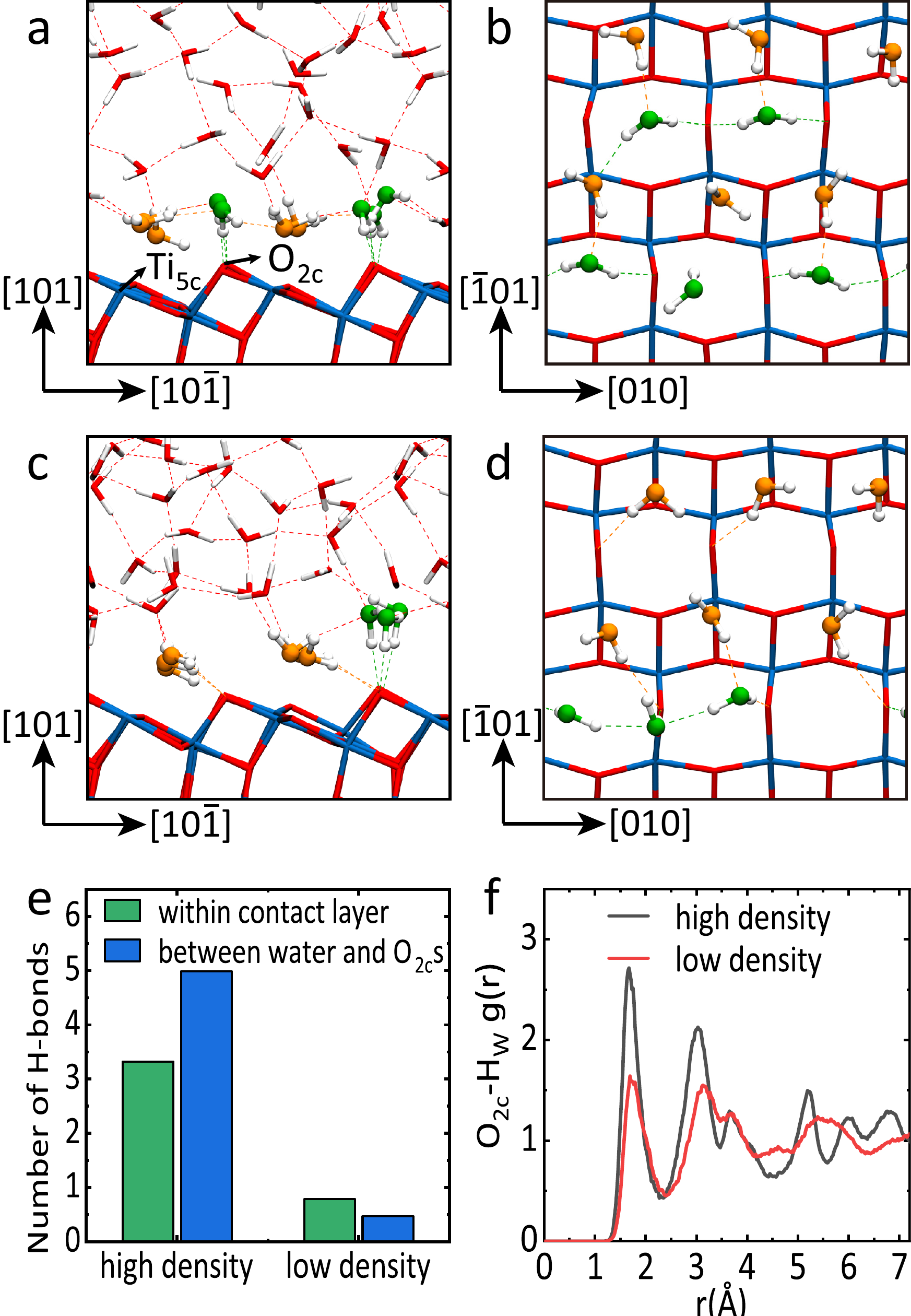}
\caption{\textbf{Comparison of the low-density and the high-density interfacial water. 
} 
(a) A side view and (b) a top view of the high-density interfacial water. The oxygen and titanium atoms are represented by red and blue, respectively. Orange and green spheres highlight water molecules in the contact layer, respectively on Ti$_{5c}$ and O$_{2c}$ sites. 
(c) A side view and (d) a top view of the low-density interfacial water. 
(e) The number of hydrogen bonds formed in two density models. Green bars correspond to hydrogen bonds between water molecules within contact layer, and blue bars correspond to hydrogen bonds formed between O$_{2c}$ and water. 
(f) The radial distribution function of O$_{2c}$-H$_{W}$ pairs, where H$_{W}$ is the hydrogen atom in liquid water. 
}
\label{fig:1} 
\end{figure}

Anatase(101) surface has a sawtooth-like structure composed of both five-fold coordinated titanium (Ti$_{5c}$) atoms and two-fold coordinated oxygen (O$_{2c}$) atoms along the [010] direction, as highlighted in Fig.\ref{fig:1}. 
Compared with their coordination number in the bulk, O$_{2c}$ and Ti$_{5c}$ are under-coordinated and are easy to capture water. 
Previous studies of water adsorption on anatase(101) surface in ultrahigh vacuum have suggested both Ti$_{5c}$ and O$_{2c}$ are favorable bonding sites, presenting two ordered layers of water. This structure of water over-layer is also reproduced in many previous simulations\cite{tiloccaDFTGGADFTSimulations2012,zhaoStructurePropertiesWater2012,aschauerInitioSimulationsStructure2015,selliWaterMultilayersTiO2017}.
For liquid water-anatase interface at room temperature, however, the structure is not  clear\cite{nadeemWaterDissociatesAqueous2018}.

Fig.\ref{fig:1}a-b shows an interfacial water structure observed in the simulation on the pristine substrate. The structure of contact water layer resembles the over-layer structure on the anatase (101) surface in vacuum, containing alternating rows of adsorbed water. 
One row is along the Ti$_{5c}$s (adsorbed water highlighted in yellow) and the other is along the O$_{2c}$s (adsorbed water highlighted in green). 
Water molecules bonded to O$_{2c}$ mainly form donating hydrogen bonds with neighboring O$_{2c}$s and accepting hydrogen bond with water molecules on Ti$_{5c}$. 
Because the structure contains two bonded water molecules per Ti site, effectively covering all surface sites, we name it as the high-density interfacial water
\footnote{The density of interfacial water can not be directly compared with bulk water because of the arbitrariness of the choice of the interfacial regime. However, for a rough comparison the lateral density of the high-density interfacial water on anatase (101) is slightly smaller than the lateral density of the basal bilayer of ice-Ih by approximately $\sim$ 10\%. The high-density interfacial water is named because it has the highest density of interfacial water observed on this surface.}.

Prior to a systematic discussion of the impact of defects on the interfacial water structure, we first discuss another typical interfacial water structure identified in our simulations. In this structure (Fig.\ref{fig:1}c-d), water molecules on the O$_{2c}$ row mostly form no more than one hydrogen bond with O$_{2c}$, in contrast to two hydrogen bonds in the high-density model. In addition, water molecules on Ti$_{5c}$ may directly form hydrogen bond with the O$_{2c}$ rather than with adsorbed water molecules (orange molecules in Fig.\ref{fig:1}a,c). In this unconventional structure, fewer water molecules feature in the contact layer, thus we refer it as the low-density interfacial water 
\footnote{The high-density and low-density water discussed here refer to interfacial water structures with relatively different lateral water density, which are different from the high-density and low-density phases in supercooled bulk water, a topic that is beyond the scope of this study.}
Fig. 1e further shows that the low-density interfacial water has fewer hydrogen bonds within the contact layer, and fewer hydrogen bonds between water and the surface.
Such differences are also shown in the radial distribution function of water hydrogen (H$_{W}$) around O$_{2c}$ (Fig.\ref{fig:1}f). In the high-density structure, more H atoms prefer to stay bonded to O$_{2c}$s, while the density of H atoms is obviously lower in the low-density model. 

\begin{figure}
\includegraphics[width=3.25 in]{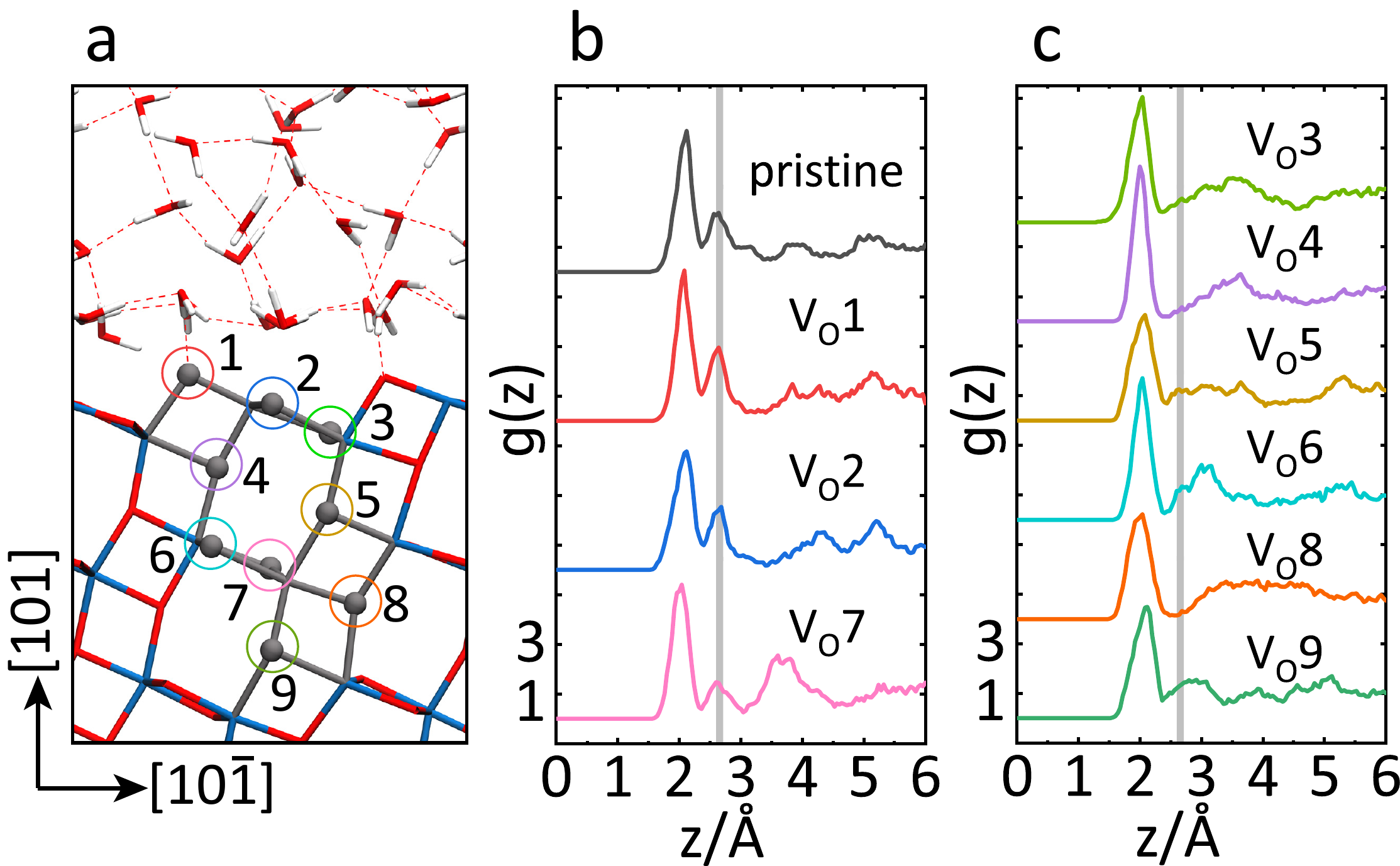}
\caption{\textbf{Structural model and g(z) profiles of ten systems.} 
(a) Side view of a snapshot from the simulation using the pristine substrate. The nine oxygen vacancy sites of defected systems are labeled with numbers 1 to 9 corresponding to the 1st to 9th oxygen layer under the surface. (b,c) g(z) profiles for all systems, which are classified as high-density interfacial water (b) and low-density interfacial water (c). The colors and labels are consistent with those in panel (a). In each g(z) profile, the reference point (z=0) is chosen as the average z value of all Ti$_{5c}$s of the whole trajectory. The vertical grey line indicates the position of the second peak on g(z) profiles of the conventional high-density interfacial water. The interfacial water density integrated from g(z) profiles are shown in Fig. \ref{fig:3}c. The convergence tests are discussed in Fig. S1, Fig. S2.
}
\label{fig:2} 
\end{figure}

Another measure to distinguish the low-density and the high-density interfacial water is the g(z) profile. 
Fig.\ref{fig:2}b,c shows all g(z) profiles of ten different simulations, including a pristine substrate and nine substrates with oxygen vacancy, which are labeled from $V_O1$ to $V_O9$ according to the depth of the oxygen vacancy. The relative stability of these 9 vacancy sites is yet to be established, therefore as a first step it is important to understand how each vacancy site impacts the interfacial water structure. In reality, vacancies may migrate across the substrates and have a dynamical influence on the interfacial water structure.
Fig.\ref{fig:2}a shows a sideview of the pristine model, and the colored circles highlight the positions of oxygen vacancy in other nine substrates. Overall from g(z) profiles, interfacial water in these ten simulations can be categorized into 4 high-density (Fig.\ref{fig:2}b) and 6 low-density (Fig.\ref{fig:2}c) structures. 
For both the high-density and the low-density interfacial water, a sharp peak at about 2 \AA\ corresponds to water adsorbed on Ti$_{5c}$. 
In the high-density model, water molecules bonded to O$_{2c}$s give rise to a second peak at about 2.6  \AA, while the peak is missing for the low-density model (grey shaded area in Fig.\ref{fig:2}b,c).
An immediate consequence of such structural changes is the density reduction of interfacial water, and a quantitative analysis will be discussed later.

\begin{figure}
\includegraphics[width=3.75in]{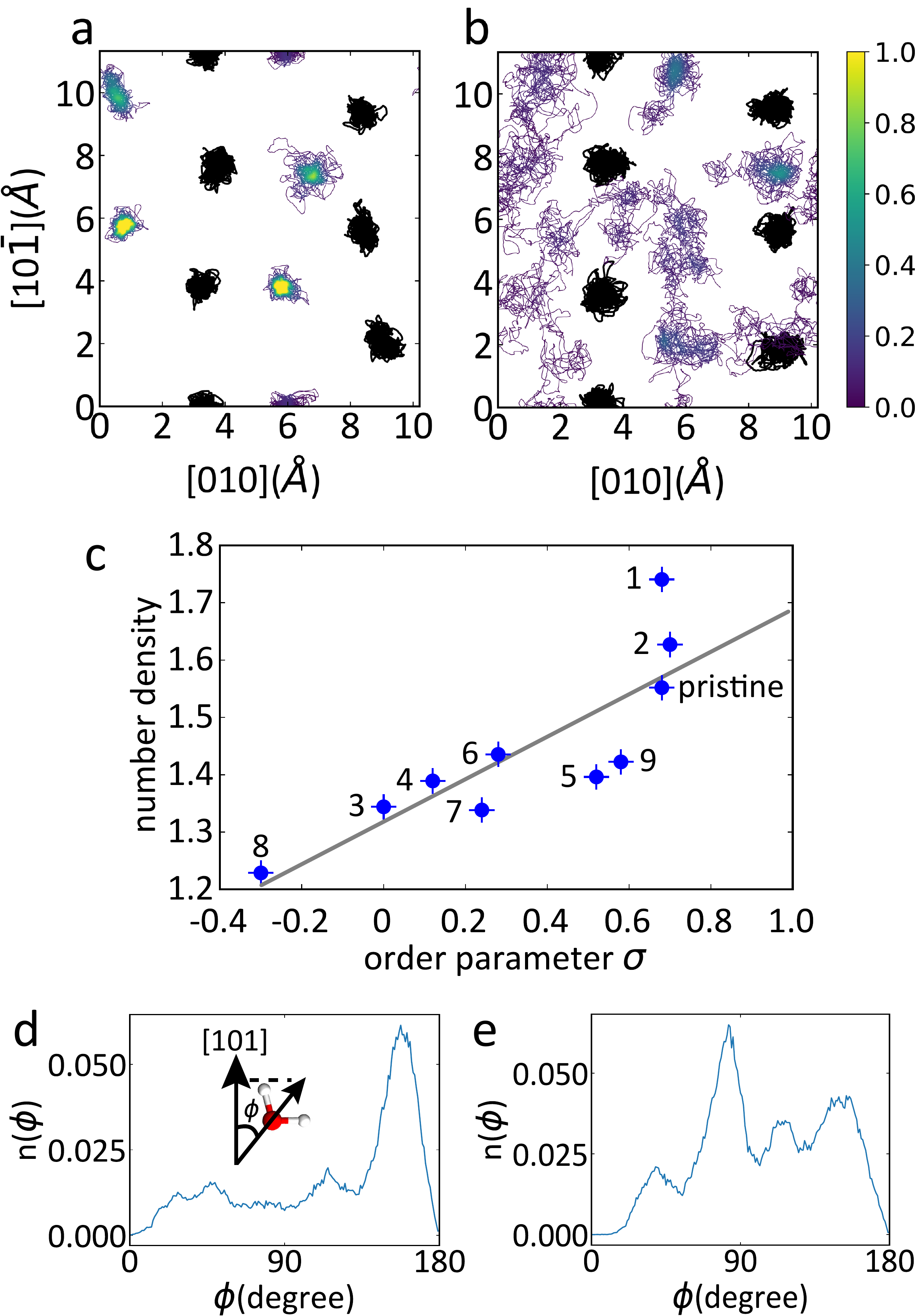}
\caption{\textbf{Fluxional properties.}
Projected trajectories for (a) The pristine system and (b) V$_O$8 systems. Fig. S7 plots projected trajectories for all ten systems. The position of water is identified by the oxygen atom. A uniform sequential colormap of viridis represents the lateral distribution density of a water molecule adsorbed on O$_{2c}$. Water molecules on Ti$_{5c}$ are plotted with black lines. Water molecules within the first shell, with a cutoff radius of 3 \AA\ from O$_{2c}$ and 2.5 \AA\ from Ti$_{5c}$, are taken into account. (c) The number density (water molecules bonded to O$_{2c}$) and order parameter of the ten systems labeled. The grey line is a guide-of-eye, showing a positive relation between the density and the ordering parameter. (d) and (e) The distribution of water orientation $\phi$ in the pristine substrate and the V$_{O}$8 system, respectively. Inset shows the definition of $\phi$, namely the angle between surface normal and the orientation of a water molecule.
}
\label{fig:3} 
\end{figure}

Besides the density reduction, the low-density interfacial water is accompanied by a fluxional feature, which are reflected in Fig.\ref{fig:3}a-b by projected trajectories of O$_W$ in x-y plane. There are six black trajectories indicating six water molecules on Ti$_{5c}$. Because water bonding energy to Ti$_{5c}$ is significantly higher than hydrogen bonding energy to O$_{2c}$\cite{doi:10.1021/jp0275544,dahalFormationMetastableWater2017}, these molecules remain adsorbed in all simulations. 
Difference appears for water molecules on O$_{2c}$s (viridis lines), whose trajectories are concentrated in the high-density model, while they are extended in the low-density model.

Another feature indicated by the trajectory projection is that the high-density interfacial water layer is ordered, whereas the fluxional interfacial water layer of the low-density model is disordered.
To quantify the ordering, we define an order parameter $\sigma$ based on the lateral angular distribution of water. The order parameter is calculated by $\sigma=\frac{1}{N}\sum_i^N\cos(3\theta)$ in polar coordinate system, where N is the total number of water molecules within a radius of 3  \AA\ from O$_{2c}$ and $\theta$ is the polar angle of the molecule (on Ti$_{5c}$)-molecule (on O$_{2c}$) line with respect to the x-axis (Fig. S4a). Orderly adsorbed water molecules on Ti$_{5c}$ and O$_{2c}$ sites feature a distribution with $\theta=0$, $\theta=\frac{2}{3}\pi$ and $\theta=\frac{4}{3}\pi$, where $\sigma$ gets a maximum value of 1. In contrast, $\sigma$ has a minimum value of -1 in directions around $\theta=\frac{1}{3}\pi$, $\theta=\pi$, and $\theta=\frac{5}{3}\pi$.
As a result, for relatively ordered distribution of water in the high-density model, $\sigma$ approaches 1, whereas the fluxional distribution in the low-density model results in a smaller, possibly negative, value of $\sigma$. 

Furthermore, Fig.\ref{fig:3}c indicates there is a relation between the density of interfacial water and the order parameter $\sigma$.
The number density of water molecules in Fig.\ref{fig:3}c is defined by the number of water molecules per unit cell within the region of 3  \AA\ above the Ti$_{5c}$ surface. 
Overall, the higher the density is the larger the order parameter $\sigma$ is, meaning the structure is more ordered.
However, when the density is lower the structure is less ordered and more fluxional.
The disordering of the contact layer is also reflected on the orientation of water molecules.
Fig.\ref{fig:3}d-e plots the distribution of molecular orientation $\phi$ of water above O$_{2c}$ rows, which are hydrogen bonded with O$_{2c}$s directly or with the water adsorbed on Ti$_{5c}$. 
For the ordered high-density model, there is a clear preference of water molecules aligning in a double-leg orientation, which peaks at about 160 degree with respect to the surface normal (Fig.\ref{fig:3}d).
In contrast, a typical fluxional interfacial model would break such a preference (Fig.\ref{fig:3}e).
This difference in orientation indicates the fluxional molecules alternately form donating hydrogen bonds with O$_{2c}$ and the bulk liquid.
The distribution profiles of all ten systems are shown in the SI (Fig. S5, Fig. S6), supporting the analyses discussed.
In addition, the fluxional behaviors are accompanied by dynamic exchanges between the contact layer and the bulk liquid (Fig. S8). 

From above results, a particularly interesting finding is that the density reduction and the fluxional interfacial water layer is observed on substrates with deep vacancies.
Even for the V$_O$7 system, showing a similar second peak on g(z) to the conventional high-density interfacial water model, other analyses of order parameter (Fig.\ref{fig:3}c), water orientation (Fig. S5) and hydrogen bonded water number (Fig. S6) clarify its fluxional nature.
V$_O$9, located $\sim$ 9 \AA~under the interfacial water layer, is the deepest vacancy considered in our simulations, and our results suggest that deeper vacancies would have similar effects.
This means that only the pristine substrate and the substrates with surface vacancies (V$_O$1 and V$_O$2) maintain the high-density interfacial water. 
Given the fact that in TiO$_2$ oxygen vacancies are ubiquitous at different depths across the substrate, low-density fluxional interfacial water may appear generally. 
The density reduction appeared in our simulation is in line with the experimental results of I.M.Nadeem et al., in which they found that a 50\% coverage of water on O$_{2c}$ fits best with measurements \cite{nadeemWaterDissociatesAqueous2018}. 

\begin{figure}
\includegraphics[width=3.75in]{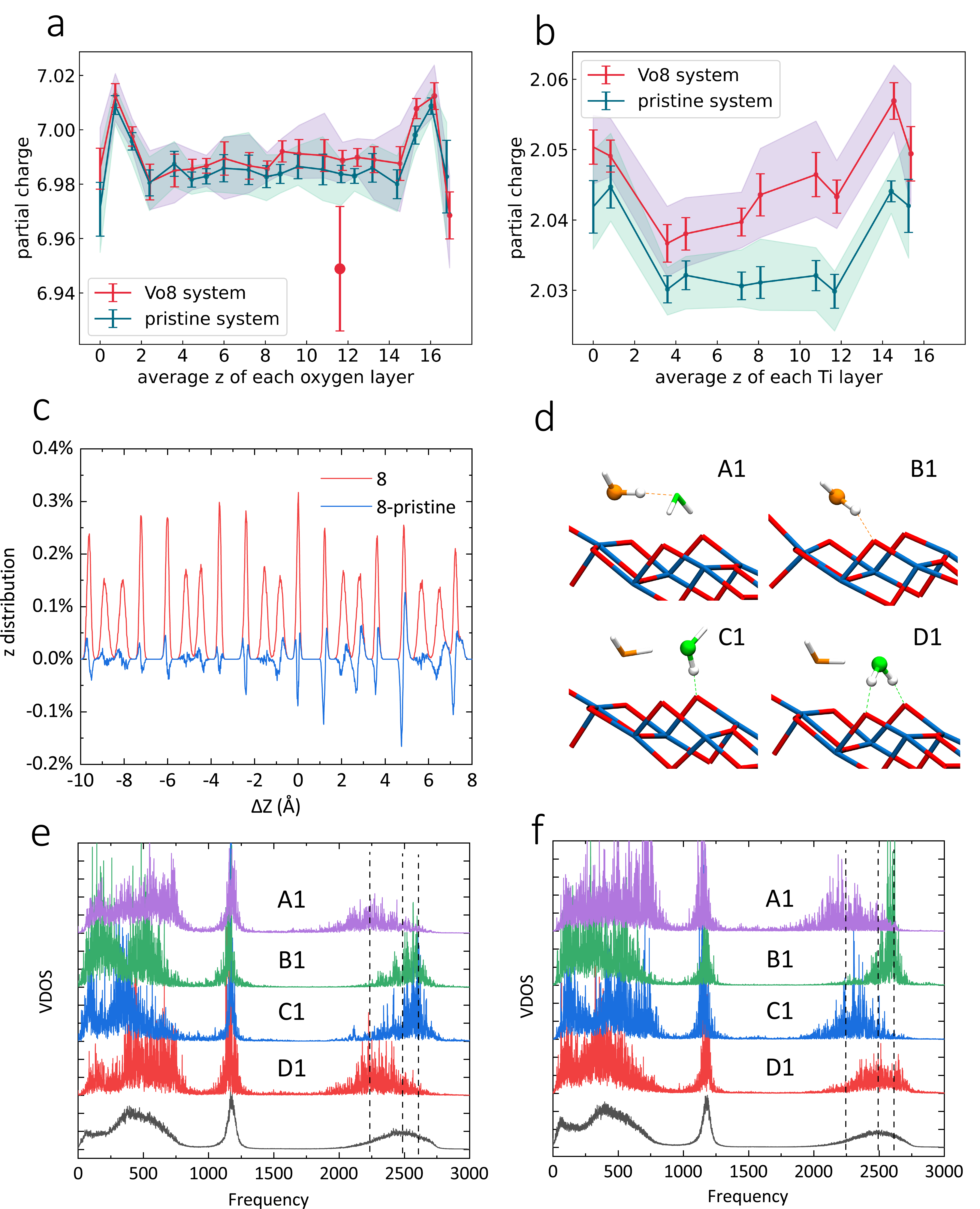}
\caption{\textbf{Substrate atomic position distribution, substrate charge distribution and VDOS of interfacial water}.
Average partial charge of every oxygen (a) and titanium (b) substrate layer. Error bar and shadow region indicate the standard deviation and interval between minimum and maximum respectively. The average is calculated every 1000 steps and every 6 atoms in a substrate layer (10 layers for Ti and 20 layers for O). The reference zero point of z coordinate is chosen to be the average z value of atoms in the lowermost interfacial layer. (c) Red line is the distribution density of oxygen atoms (20 layers) except for the dangling oxygen atom in the V$_{O}$8 system. Blue line is the difference from the same distribution density in pristine system, showing the difference in two substrates are extended across the whole substrate. $\Delta$Z is defined as the difference between z value of an oxygen atom and the average z value of all six oxygen atoms in 9th layer. The 9th oxygen layer is chosen as the reference for the plot, since its average oxygen position is unaffected by the existence of V$_O$. (d) Snapshots (A1-D1) show the structures concerned in vibrational density of state (VDOS) of pristine system (e) and of V$_{O}$8 system (f). Black lines in VDOS represent the liquid reference in their systems.}
\label{fig:4} 
\end{figure}

It is also worth noting that our results do not contradict with previous studies where water dissociation was discussed. Recently, Calegari Andrade et al. investigated the spontaneous dissociation of water on TiO$_2$ surface using machine learning force fields and suggested that spontaneous water dissociation requires nanoseconds for equilibrium\cite{andrade_free_2020}. Here we focus on the interfacial structure of water, characterized by the distribution of oxygen positions, whose equilibrium is often reached within a few tens of picoseconds. In fact, water dissociation induced by surface V$_O$ is also observed in our simulations with V$_O$1 and V$_O$2. Dissociation, however, does not lead to the density reduction of interfacial water. The fact that surface vacancies (V$_O$1 and V$_O$2) have minor effects on the interfacial water structure is rather counter-intuitive because surface oxygen vacancies are reactive to water, and lead to the formation of hydroxide groups on the surface.

Our counter-intuitive results suggest there is very delicate balance of competing interactions at the interface associated with the details of substrate. To provide deeper physical insights, we further compared the position and charge distribution of the V$_O$8 system and the pristine substrate. Fig.\ref{fig:4}a-b plots the Bader charge of oxygen and titanium atoms on each layer across the pristine and V$_O$8 substrates. In comparison to the pristine system, oxygen vacancies result in excess electrons in the substrate. We note that although the difference is a few percent, this magnitude of change is large enough to induce an electrostatic energy difference of tens of meV simply based on electrostatics, which is comparable to the 330K temperature (28 meV). Such a difference is enough to tip the balance of the competition between hydrogen bonding network and adsorption on bridging oxygen, thus leading to a new interfacial water structure on TiO$_2$. For example, previous studies have shown a competition between two water molecules for hydrogen  bonding to the bridging oxygen that raises up the adsorption energy by about 33 meV\cite{heLocalOrderingElectronic2009}. Slight electron deficit on bridging oxygen (Fig.\ref{fig:4}a) in defected system leads to weaker hydrogen bonding, hence pushing one of the competing water molecules away from the bridging oxygen.

Previous studies have shown that excess electrons can migrate from subsurface to the surface, affecting hydrogen bonding, adsorption and dissociation of water\cite{selcukFacetdependentTrappingDynamics2016,chenSmallPolaronsJanus2020,yimVisualizationWaterInducedSurface2018}. We note that the behavior of excess electrons is quite sensitive to DFT methods used. For example, excess electrons may localize and show polaronic behavior when DFT+U methods are used. We have also carried out simulations with SCAN-rvv10 and PBE+U, which show consistent results that density reduction can be induced by deep oxygen vacancies (Fig. S3, Fig. S7-8, Fig. S10). These additional tests suggest the low-density fluxional interfacial water is not an artifact and not sensitive to the description of localized electrons and the extent of water dissociation.

Fig.\ref{fig:4}c plots the oxygen atom distribution and their difference, which shows the interfacial water structure on these two substrates are quite different, displaying a collective variation of atomic positions across the substrate layers. The collective variation is a typical elastic effect that can physically transfer interactions to a long range. Similar effects have been observed in rutile, where water adsorption and dissociation energies vary along with the substrate positions\cite{liuStructureDynamicsLiquid2010}. In the SI, we further compare the V$_O$5 and V$_O$8 systems. These two systems have oxygen vacancies located on the opposite side of a void regime, and during the simulations the oxygen atom on the opposite side vibrates in the middle, resulting in two very similar substrates structurally (Fig. S11) and electronically (Fig. S12). Overall, here we propose that the long-range effects to the interfacial water structure may entangle with minuscule atomic position and charge distribution charges of the substrate, which should be paid attention to and further investigated in the future.

Before concluding, we note that the delicate nature indicates that experimental identification of the interfacial water structure on anatase(101)-water interface is of great challenge, requiring not only advanced characterization techniques but also clean and well controlled substrate samples. The interfacial water density change reflects to some extent on the hydrogen bonding change at the interface, which can be detected experimentally using vibrational spectroscopies. Fig.\ref{fig:4}e-f shows the vibrational density of state (VDOS) from AIMD trajectories of the pristine and V$_{O}$8 systems. VDOS profiles A1-D1 in Fig.\ref{fig:4}e-f correspond to the donor O-H stretching highlighted in Fig.\ref{fig:4}d. For comparison VDOS from water molecules are shown in black lines. The red and blue shifts in the stretching mode indicate strong and weak hydrogen bonds. The strength difference in C1 and D1 between pristine and defected systems leads to the difference in the interfacial structure. Specifically, oxygen vacancy shifts the preference of local hydrogen bonding structure from D1 to C1, which is also highlighted in the dipole orientation profiles (Fig. 3d-e).

Finally, to examine whether our findings can extend to other systems, we have considered a few other substrates, namely rutile(110), anatase(001), and silica(0001) (Fig. S13, Fig. S14). These substrates cover three different kinds: (i) on rutile(110) water remains intact; (ii) on anatase(001) water dissociates; and on silica(0001) the surface is readily hydroxylated. 
In the end, on these substrates tested we do not see strong influences of deep oxygen vacancies on the interfacial water structure.
Therefore, whether there are other substrates that possess similar interesting behaviors identified on anatase(101) remains an open question.

In summary, we have reported \textit{ab initio} simulations of water-anatase(101) interface, where we have identified the low-density fluxional interfacial water due to oxygen vacancies in the substrate. 
We show that the influence of oxygen vacancies can be rather long-range and indirect, which lead to the finding that even the deepest vacancy considered in our simulations can strongly impact the interfacial water structure.
We also reveal the density reduction and induced flexibility is closely related, and
the behaviors are likely to be physically amplified by minor changes in the substrate due to the presence of oxygen vacancy.
The non-trivial long range effect identified in our study is another highlight of the complexity of interfacial water structure, and it is also potentially an opportunity for materials engineering to manipulate interfacial water properties by controlling the inner part of materials.
%

\section*{acknowledgements}

The authors thank Angelos Michaelides for helpful discussions.
This work was supported by the National Key R\&D Program of China under Grant No.2016YFA030091, the National Natural Science Foundation of China under Grant No. 11974024, and the Strategic Priority Research Program of Chinese Academy of Sciences under Grant No. XDB33000000.
We are grateful for computational resources provided by Peking University, the TianHe-1A supercomputer, Shanghai Supercomputer Center, and Songshan Lake Materials Lab.





%

\end{document}